# Molecular ordering of nematic liquid crystals in tubular nanopores: Tailoring of optical anisotropy at the nanoscale by polymer pore-surface grafting

Andriy V. Kityk
Faculty of Electrical Engineering
Czestochowa University of Technology
Czestochowa, Poland
andriy.kityk@univie.ac.at

Anatoliy Andrushchak
Institute of Telecommunications,
Radio Electronics and Electronic Technics
Lviv National Polytechnic University
Lviv, Ukraine
anat@polynet.lviv.ua

Robert Wielgosz[1], Olha Kityk[2]
Energia Oze Sp. z o.o.
Konopiska, Poland
[1]robert.wielgosz@interia.eu
[2]biuro@odnawialnaenergia.pl

Patrick Huber[1], Kathrin Sentker[2]
Institute of Materials Physics and Technology
Hamburg University of Technology
Hamburg, Germany
[1]patrick.huber@tuhh.de
[2]kathrin.sentker@tuhh.de

Przemysław Kula[1], Wiktor Piecek[2]
Faculty of Advanced Technologies and Chemistry
Military University of Technology
Warszawa, Poland
[1]przemyslaw.kula@wat.edu.pl
[2]wiktor.piecek@wat.edu.pl

Petra Goering[1], Monika Lelonek[2]
SmartMembranes GmbH
Halle, Germany
[1]petra.goering@smartmembranes.de
[2]monika.lelonek@smartmembranes.de

*Abstract* — The optical polarimetry technique is used to explore molecular ordering of nematic liquid crystals 5FPFFPP3 embedded in parallel cylindrical channels of alumina membranes with pore walls covered by a SE-1211 polymer film enhancing normal anchoring. A chemical film deposition from organic solution apparently results in inhomogeneous coverage with local islands on the channels walls. Accordingly, a coexistence of differently ordered LC regions (nanodomains), characterized by positive and negative anisotropy, results in a weak effective (average) optical birefringence. Repeated polymer deposition reduces the size of uncovered pore walls regions and thus leads to a strengthening of negative optical anisotropy observed in the experiments.

## I. INTRODUCTION

The optical anisotropy of liquid crystal (LC) based nanocomposites is defined by the molecular ordering of the LC molecules inside the tubular nanopores. Cooperative ordering of the nanoconfined LC phases results in macroscopic birefringence which in most cases substantially dominates a geometrical birefringence caused by an elongated pore geometry. This effect has been demonstrated in numerous experimental studies performed, particularly, on nanocomposites with rod-like [1,2] and disc-like [3,4] molecules of achiral nematics as well as their chiral counterparts, cholesterics [5] and ferroelectric LC materials [6], representing guest component in host alumina ($p$Al$_2$O$_3$) or silica ($p$SiO$_2$) nanoporous templates. Accordingly, the LC based nanocomposites with tailored optical anisotropy continue to be an interesting problem for both basic science and for technological reasons. Interest in such materials is also enhanced by the fact that confined LC states can be operated by external electric or magnetic fields, i.e. the effective anisotropy can be controlled. Due to this reason recent trends in electronics and optoelectronics shift many applications to a nanoscale level where nanocomposites with tailored properties appear to be in priority positions.

The molecular ordering of LC molecules in nanopores is considerably affected by interfacial interactions. Quenched disorder and geometrical confinement effects, on the other hand, have crucial influence on both their static and dynamic properties as has been demonstrated in recent optical [1,2] and dielectric studies [7,8]. Indeed a competition of these effects define a confined molecular ordering configuration. For instance, in the case of rod-like molecules a curved pore geometry plays a role of an ordering field, as introduced for the first time by Kutnjak, Kralj, Lahajnar and Žumer in their phenomenological approach (so-called KKLZ theory) [9,10]. Such a geometric field, the strength of which scales as $1/D$ (where $D$ is the pore diameter), forces to orient elongated nematic molecules along the pore axis. In the case of tangential anchoring rod-like molecules of the confined nematic phase appear to be oriented along the pore axis forming thus an anisotropic medium with *positive* birefringence. It is characterized by a *prolate* optical indicatrix with the long axis parallel to the pore channels. In contrast, for confined nematics with disc-like molecules oriented perpendicularly to the long pore axis the molecular configuration leads to an *oblate* anisotropy with a *negative* birefringence. In practice, however, nanoconfined discotic nematics usually exhibit a so-called face-on orientation [3,4] or even may form closed circularly shaped columns [11]. In both cases a positive anisotropy results. In principle, a negative anisotropy is also expected for confined calamitic nematics with dominating normal anchoring at pore walls which may result to radial molecular ordering, such as e.g. escaped radial configuration [12]. However, our previous studies on nanoconfined achiral calamatic LCs $n$CB ($n$ = 5-7) evidently shows that a parallel configuration remains energetically still more preferable. The reason for this may be that a lowering of the free energy due to normal anchoring is compensated by a positive energy contribution due to splay deformation which especially rises in pores of small diameters. Perhaps due to this reason metastable radial ordering have been observed in the pores of large diameters [12]. Negative birefringence has been noticed recently also in chiral LC materials, particularly cholesteric LC CE6 embedded into $p$Al$_2$O$_3$ membranes with SE1211-polymer treated surface walls [5]. Chiral nematics, however, differ considerably from their achiral counterparts since confined cholesterics form helical structures requiring specific consideration.

In the present work we demonstrate tunable optical anisotropy in LC-$p$Al$_2$O$_3$ nanocomposite provided by SE-1211 polymer treatment of nanochannels walls. Molecular ordering is analyzed by the optical polarimetry technique. We show that multilayer polymer coverages provide substantial modification of the optical anisotropy changing it from positive to negative one.

## II. EXPERIMENTAL

Fig.1 shows the polarimetry setup. The optical retardation of polarized light components is modulated by a photoelastic modulator (PEM, Hinds Instruments Inc.) and is modified by the static component $\Delta$, introduced by the measured sample. An interference of the polarized components on analyser (A) results in a modulated light intensity detected by the photodetector (PD). It is subsequently analysed by two pairs of lock-in amplifiers (Standford Research System, SR-830) which measure amplitudes of the first ($I_\Omega$) and second ($I_{2\Omega}$) harmonics. The acquired data are transferred via GPIB to PC for their saving and further processing. The measured optical retardation $\Delta$ is determined using the equation inserted in Fig.1, where $k$ is the factor defined by the ratio of the Bessel function at the retardation amplitude ($\delta_0=0.383\lambda$) corrected by the PD frequency response. To probe the optical anisotropy the sample in the optical temperature cell is tilted out of the normal position with a certain angle, α.

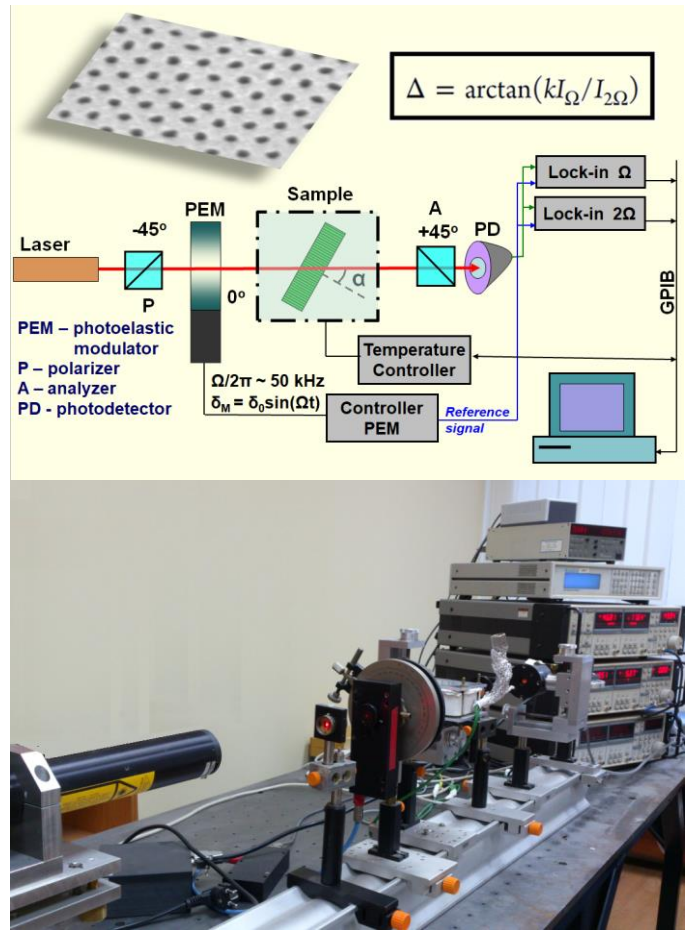

Fig.1. Picture (lower panel) and schematics setup (upper panel) of the optical polarimetry experiment.

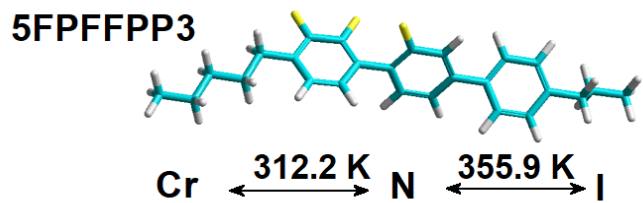

Fig.2. Schematic sketch of the nematogenic molecule(5FPFFPP3) and characteristic phase transition temperatures, I - isotropic phase, N – nematic phase, Cr – crystalline phase.

We report here optical polarimetry experiments on nonchiral nematic LC 2,3,2'-trifluoro-4-pentyl-4''-propyl-p-terphenyl (abbreviated as 5FPFFPP3) embedded into parallel arrays of cylindrical channels of alumina membranes (pore diameter 40 nm). The molecular structures of 5FPFFPP3 is presented in Fig.2.

The induced optical anisotropy may be characterised by the excess optical retardation measured at a certain constant incident angle, $\alpha$. For rodlike nematic molecules it is proportional to the orientational order parameter $Q = 0.5 \cdot \langle 3 \cdot \cos^2\varphi - 1 \rangle$, where $\varphi$ is the angle between the long axis of the molecules and the direction of preferred orientation (director), i.e. in many cases it may be used for a characterization of molecular ordering as well.

Our experiments have been performed on nanocomposites made of alumina nanoporous membranes from SmartMembranes GmbH ($D$=45 nm) with pore walls subsequently covered by SE-1211 polymer film and filled by the nematic LC 5FPFFPP3. Polymer coverage (up to three layers) have been deposited from weakly concentrated solution of SE-1211 (0.7 wt.%) in organic solvent (N,N-Dimethylformamide). Alumina membranes have been previously annealed at 180 $^{o}$C for about 1 hour and then immersed into the organic polymer solution for 24 hours. After that they were subsequently dried at 25 $^{o}$C (~ 2 hours) and 70 $^{o}$C (~2 hours) and finally polymerized at about 170 $^{o}$C for 1/2 hour. For the multilayer coverage the procedure mentioned above has been repeated a number of times.

III. EXPERIMENTAL AND DISCUSSION

Fig.3 shows temperature dependences of the optical retardation $\Delta(T)$ of nematic LC 5FPFFPP3 embedded into parallel-arrays of cylindrical channels of alumina membranes ($D$=45 nm) with pore walls covered by SE-1211 polymer nanometric layer(s), see labelled. The gray broken line [Section (a)] indicates the reference behavior of the 5FPFFPP3-$p$Al$_2$O$_3$ nanocomposite recorded for the untreated alumina matrix. For such types of nanocomposites the retardation increases continuously upon cooling indicating thus a preferred alignment of nematic molecules parallel to the channel axes. The observed behaviour suggests that native silica or alumina pore walls exhibit tangential anchoring resulting in positive birefringence. Interfacial interactions, on the other hand, result in a continuous temperature evolution of the measured retardation with residual birefringence extending far above the paranematic (PN)-to-nematic(N) transition temperature. These characteristics fit well with the KKLZ model for confined nematic LCs [9,10].

Polymer SE-1211 enhances normal anchoring thus using it as wall coverages should modify the optical anisotropy of LC nanocomposite materials. Our experimental studies show that the temperature dependences of the optical retardation are not reproducible in subsequent cooling-heating cycles and the thermal history depends sensitively on the number of polymer coverages being deposited. With a single layer polymer coverage [Fig.3, section (a)] $\Delta(T)$ behaves in the first temperature circle in a manner similar to nanocomposites based on untreated alumina membranes. However, in subsequent temperature circles evident deviations from such behaviour is observed. In the confined N phase the optical retardation for subsequent temperature circles decreases by more than 40%. Nevertheless, it remains still positive even after a large number of measuring cycles, i.e. the further behaviour exhibits a saturation character. Subsequent polymer coverages result in a further decreasing of the optical birefringence. For the two layer coverage [Fig.3, section (b)] we find practically a birefringence magnitude of zero already at the 1st heating run. Subsequent cooling-heating runs result in a more pronounced negative optical birefringence.

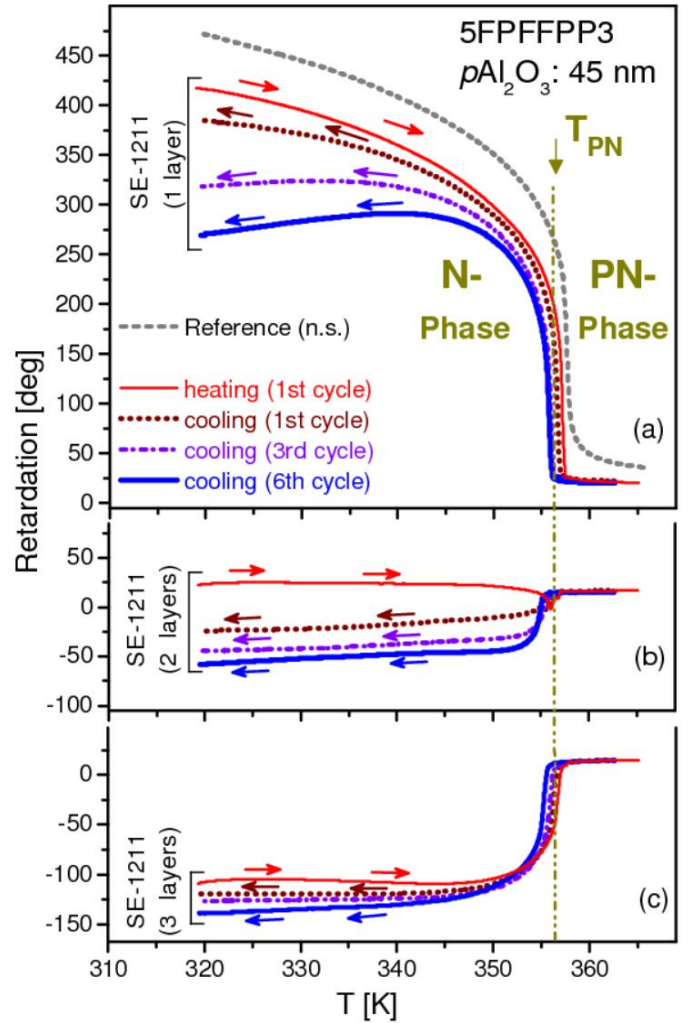

Fig.3. Temperature dependences of the optical retardation, $R$, of nematic LC 5FPFFPP3 embedded into parallel-arrays of cylindrical channels of alumina membranes ($D$=45 nm) with pore walls covered by SE-1211 polymer nanometric layer(s), incident angle, $\alpha$ = 36$^{o}$. Panel (a) (b) and (c) depict $R(T)$-dependences for the nanocomposites consisting of polymer treated alumina membranes with pore wall surface graftings resulting from one, two and three subsequent polymer coverages, respectively. Measurements have been performed for a sequence of heating-cooling cycles, as marked by arrows. Note that for clarity not all temperature cycles are shown, see labels. The gray dashed line [Section (a)] indicates the reference behavior of the 5FPFFPP3-$p$Al$_2$O$_3$ nanocomposite with untreated alumina matrix.

Three layer coverage [Fig.3, section (c)] makes these changes even more evident. Already in the 1st cooling-heating run we find a substantially negative birefringence dominated presumably by normal anchoring at the pore wall. After 15-20 temperature cycles the measured absolute value of the negative retardation saturates at a magnitude of about 40% of its reference value for parallel axial configuration, i.e. the magnitude of the saturated positive retardation measured in nanocomposite with untreated alumina membranes. It is just a bit less of the corresponding theoretical limit value (~50%) expected for this case.

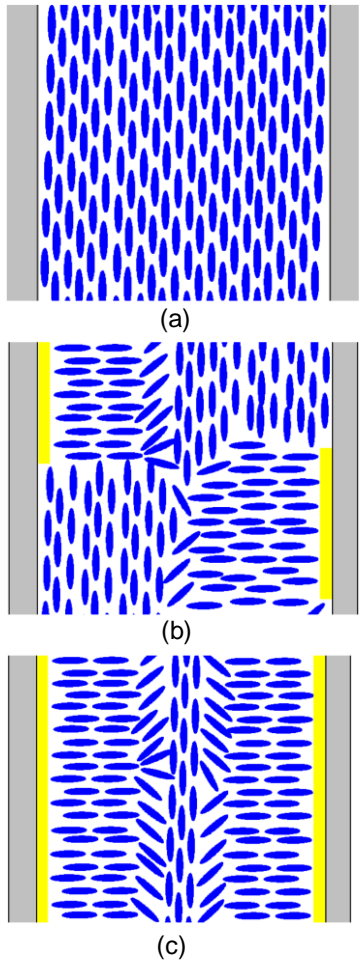

Fig.4. Sketch of the molecular ordering of nematic LC molecules in a nanochannel. Section (a): Untreated pore walls. The native $p$Al$_2$O$_3$ surface enforces tangential anchoring with nematic ordering parallel to the long channel axis resulting thus in a nanocomposite with positive optical anisotropy. Section (b): $p$Al$_2$O$_3$ surface with inhomogeneous coverage by a SE-1211 polymer film (yellow) resulting in a coexistence of ordered regions (domains) with positive and negative anisotropy. Section (c): $p$Al$_2$O$_3$ surface with homogeneous coverage by a SE-1211 polymer film resulting in an escaped radial structure with strongly dominating negative optical anisotropy.

Our interpretation of the obtained results are illustrated in the schematic sketches of molecular ordering in Fig.4. Native alumina surfaces enforce tangential anchoring which in combination with the geometric ordering field provided by the channel wall curvature results in stable parallel axial configuration, see Fig.4(a). A polymer SE-1211 nanometric coverage is expected to enforce normal anchoring. However single and even double coverages apparently does not result in a homogeneous polymer film. Chemical deposition presumably leads to local islands as it is schematically demonstrated in Fig. 4(b). Accordingly, a coexistence of differently ordered regions, i.e. nanodomains with positive and negative anisotropy is expected. Particularly, the weak optical anisotropy in Fig.3(b) may be explained assuming that the contributions from domains with positive anisotropy (tangential anchoring) and domains with negative anisotropy (normal anchoring) compensate each other. We suppose that repeated deposition reduces uncovered pore walls regions. This explains a subsequent strengthening of negative anisotropy observed in the experiments. In the full-coverage limit [Fig.4(c)] an escaped radial structure with dominating negative anisotropy is expected.

## IV. CONCLUSIONS

In the present work we explore optical anisotropy of LC based nanocomposites made of porous alumina membranes with channel walls treated by SE-1211 polymer. The optical anisotropy is analyzed by the optical polarimetry technique. We demonstrate that multilayer polymer coverages provide substantial modification of the optical anisotropy changing it from a positive to a negative one. Native alumina pore walls provide tangential anchoring which in combination with geometric ordering fields, caused by channel wall curvature, results in a stable parallel axial configuration with strong positive birefringence. A polymeric SE-1211 nanometric coverage, in contrast, enhances normal anchoring. The experiments indicate that chemical film deposition from organic solution apparently results in inhomogeneous coverage with local islands on the channels walls. Accordingly, a coexistence of differently ordered LC regions characterized by positive and negative anisotropy and weak effective (average) optical anisotropy is found. Repeated deposition reduces uncovered pore wall regions. This explains a subsequent strengthening of negative anisotropy observed in the experiment. Nanoconfined LCs with nanodomain substructure represent extraordinary systems which may be of interest for parametrical optical effects such as e.g. electrooptics or magnetooptics. The existence of domains results often in a large response with respect to applied external fields, particularly because of domain wall mechanisms, see e.g. the mechanisms considered in Ref.[13]. For the future alternative guest materials may be a series of organic crystals (e.g. biphenyl, naphthalene or benzil), as well as water soluble ferroelectric crystals, like e.g. TGS [14,15], KDP [16] or GPI [17] or proper incommensurate ferroelastics of the Cs$_2$XY$_4$ group [18]. The exploration of the orientation of the nanocrystals inside the nanopores depending on the polymer pore wall grafting in combination with studies on the optical anisotropy, optical parametric and nonlinear optical effects would be of particular scientific [19] and technological interest.

ACKNOWLEDGMENT

This work is a part of a project that has received funding from the European Union's Horizon 2020 research and innovation program under the Marie Sklodowska-Curie grant agreement No 778156. K. S. and P. H. profited from the support within the Collaborative Research Initiative SFB 986, Tailor-Made Multi-Scale Materials Systems, projects B7, Z3, Hamburg (Germany) funded by the Deutsche Forschungsgemeinschaft (DFG).